\documentclass{aa}
\usepackage{graphics,psfig}
\begin{document}

   \thesaurus{1                  
	       (19.09.1;19.19.2)}
   \title{New shell radio supernova remnant G16.2$-$2.7}

   \author{Sergei A. Trushkin}

   \offprints{S. A. Trushkin}

   \institute{Special astrophysical observatory,
	      Niznij Arkhyz, 357147 Russia\\
	      email: satr@sao.ru
             }

   \date{Received 25 October 1999 / accepted  1  November 1999}

   \maketitle

   \begin{abstract}
     The extended radio source G16.2$-$2.7 is detected as a
     new previously uncataloged Galactic supernova remnant.
     Its non-thermal radio spectrum has spectral index $\alpha=-0.51$,
     with S$_\nu$(1\,GHz) = 2.08 Jy, as being measured with the RATAN-600
     radio telescope.
     The NRAO VLA Sky Survey (NVSS) map at 1.4 GHz shows a shell-like
     bilateral structure. The similar smoothed image from
     the Effelsberg survey at 2.7 GHz is discussed.
     The angular diameter 17\arcmin\ of a circular shell is fitted
     to brightness peaks meanwhile the outer diameter
     D$_{\mathrm{max}}$ = $18\farcm4$  and the width $\Delta$R=$1\arcmin$
     are fitted with the model of a spherical optically thin hollow shell.
     The surface brightness of G16.2$-$2.7:
     $\Sigma$(1GHz)$=(1\pm0.1)10^{-21}$W\,Hz$^{-1}$m$^{-2}$sr$^{-1}$.
     The peaks in the shell arcs are highly polarized at 1.4 GHz.

   \end{abstract}

%

\section{Introduction}

The total number of supernova remnants (SNRs) is estimated by different
methods, and it is generally accepted to be about 300--1000 detectable SNRs
in the Galaxy.
Green's (\cite{Green}) catalog includes 220 confirmed SNRs
and several dozens of possible or probable ones. Most of them are radio SNRs.
Thus a search for new SNRs is an important task of observational
radio astronomy. Such searches have been made by Whiteoak \& Green (\cite{WG92}),
Gray (\cite{GraySNR,GrayNew}), Duncan et al. (\cite{Dun97}).
Using the radio morphology, Weiler (\cite{W83}) divided the Galactic
SNRs on three classes of shell-like, crab-like or plerionic and
mixed or composite ones.
As a rule, in these searches the shell or composite SNRs have been found, and
the shells dominate the total sample.

Trushkin (\cite{Sur96,Sur98}) searched for new SNRs in the Galactic plane
survey with the RATAN-600 radio telescope in the First and Fourth
Galactic quadrants between $l = 343\degr$ and $l = 19\degr$ and
$|b| < 5.5\degr$ at 0.96, 3.9 and 11.2 GHz.
A dozen extended non-thermal sources which could be new SNRs
have been revealed.
A analysis of the NRAO VLA Sky Survey (NVSS) maps (Condon et al.
\cite{NVSS}) is extremely helpful for such a search in the Galactic plane.
In spite of the apparent insensitivity of NVSS to extended radiation,
nearly 80 known Galactic SNRs are visible in NVSS maps.
Trushkin (\cite{Atlas}) created the collection of the SNR images that is
released via the World Wide Web.
This induces us to search for new shell SNRs in the original
$4\degr\times4\degr$ NVSS maps. Excluding known HII regions
or planetary nebulae, we have selected nearly 20 shell SNR candidates.
One of them, G16.2$-$2.7, shows all the necessary properties to recognize
it as a new SNR.

\begin{figure}
\vspace{0cm}
\vbox{\psfig{figure=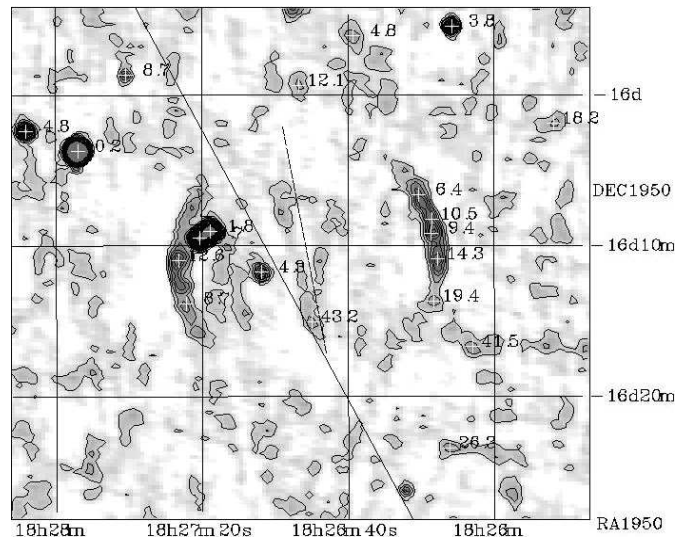,width=8.8cm}}
\vspace{0cm}
\caption[]{%
Radio map of the new SNR G16.2-2.7 from the NVSS at 1.4 GHz.
Contours are with steps 1 mJy/beam from the first contour 0.7 mJy/beam,
where beam is equal to $45\arcsec\times45\arcsec$. The NVSS sources from
Table \ref{src} are marked by white crosses and numbers annotate the power
of polarization in per cent.
The solid inclined line represents a Galactic latitude $b=-2\degr40\arcmin$.
The dashed line represents the axis of symmetry for this SNRs (see text).
The rms in the regions out of it: 1$\sigma\approx0.6$ mJy/beam
and the mean rms in NVSS maps: 1$\sigma=0.45$ mJy/beam}.
\label{map}
\end{figure}

\section{The radio maps}

In Fig.\ref{map} the NVSS-map of G16.2$-$2.7 at 1.4 GHz is shown.
Here the maps are plotted with the ``Skyview'' package
(Ebert et al. \cite{sky}).
The circular shell structure of the SNR has the angular diameter $17.0\pm0.2$\arcmin,
while the integral flux density in the map, S$_\nu = 0.7\pm0.3$ Jy, is much
lower than the value extrapolated from the  spectrum (see below).
The large uncertainty of the NVSS flux depends strongly on the background
level definition and could only be a lower limit because
the NVSS suffers from lack of zero-spacing data and
its images are insensitive to smooth radio structures much larger
than several arcmin.

The flux weighted centroid of the source has the
Galactic coordinates: $l$=16\fdg167 and $b=-2$\fdg689 or equatorial ones:
RADEC1950 = $16^h27^m08^s$, $-$16\degr09\arcmin50\arcsec.

Recently  Gaensler (\cite{bilat}) has investigated the nature of the
the bilateral SNRs, one of them, G03.8-0.3, has a bilateral structure
very similar with G16.2$-$2.7.
Their surface brightness, angular sizes are close. We could estimate
the value of $\psi$ defined to be the acute angle between the symmetry axis
of the SNR and the Galactic plane. We fit the symmetry axis using only
the bright circular arcs.  This gives a value $\psi=17\degr\pm3\degr$.
Thus the symmetry axis is aligned close to the Galactic plane (see Fig.\ref{map}).
It is not clear whether the central weak filament with a brightness of nearly
1.5 mJy/beam ($\sim3-4\sigma$) located close
to the symmetry axis is real and associated with this SNR or not.
The new radio mapping are needed.

\begin{table}
\caption{NVSS sources around of the SNR G16.2$-$2.7}
\begin{tabular}{rccrrl}
\hline
N  &RA1950     &  DEC1950  &S$_\nu$& P$_{pol}$& In \\
   & hhmmss.ss &~ddmmss.s  & (mJy) &  \%      & SNR?\\
\hline
1  & 182543.88 &$-$160153.1 &  3.3 & 18.2& no  \\
2  & 182605.89 &$-$161642.6 &  4.1 & 41.5& no  \\
3  & 182611.71 &$-$155524.8 & 15.6 &  3.8& no  \\
4  & 182612.16 &$-$162325.8 &  3.8 & 26.3& no? \\
5  & 182615.50 &$-$161052.2 & 18.2 & 14.3& yes \\
6  & 182616.68 &$-$161340.3 &  3.6 & 19.4& yes \\
7  & 182617.19 &$-$160814.8 & 15.3 & 10.5& yes \\
8  & 182617.36 &$-$160909.9 &  5.3 &  9.4& yes \\
9  & 182620.78 &$-$160636.4 & 14.0 &  6.4& yes \\
10 & 182638.72 &$-$155603.8 &  4.2 &  4.8& yes \\
11 & 182649.85 &$-$161502.8 &  3.7 & 43.2& yes?\\
12 & 182653.45 &$-$155920.6 &  3.3 & 12.1& yes?\\
13 & 182703.77 &$-$161145.8 &  9.3 &  4.3& yes?\\
14 & 182717.80 &$-$160900.9 & 80.0 &  1.8& no  \\
15 & 182720.69 &$-$160930.0 & 53.4 &  0.7& no  \\
16 & 182724.39 &$-$161349.1 & 12.6 &  8.7& yes \\
17 & 182726.43 &$-$161058.1 & 25.3 & 12.6& yes \\
18 & 182740.81 &$-$155839.8 &  4.6 &  8.7& no  \\
19 & 182754.01 &$-$160341.9 &352.9 &  0.2& no  \\
20 & 182808.20 &$-$160221.5 & 18.9 &  4.8& no  \\
\hline
\end{tabular}
\label{src}
\end{table}

\begin{figure}
\vspace{0cm}
\vbox{\psfig{figure=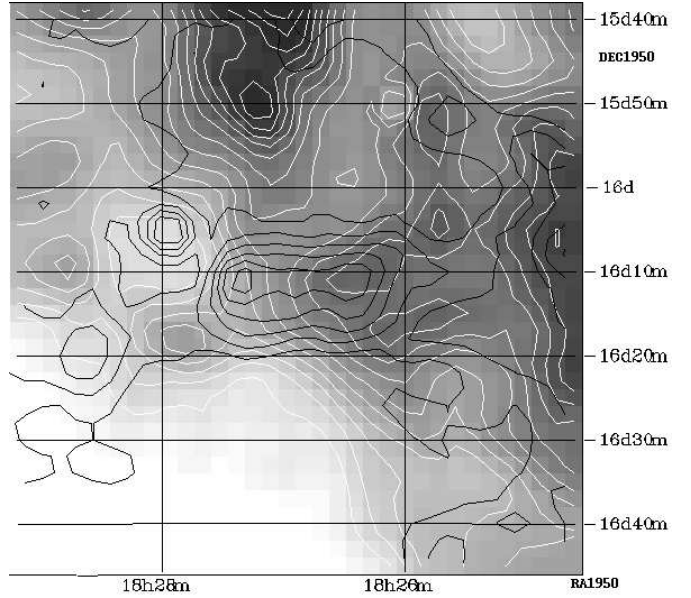,width=8.8cm}}
\vspace{0cm}
\caption[]{%
The grey scale ISSA (IRAS) $1\fdg1\times1\fdg1$ map
at 60 $\mu$m superposed on contour 2.7 GHz intensity map (RFRR).
The black contour levels are drawn linearly with steps of 75 mK from
the first contour level 160 mK of T$_B$. The grey-scale and white
contours are from 70 MJy/sr with step 4 MJy/sr. The pixel of the both
maps is equal $2\arcmin$.
}
\label{ir}
\end{figure}

\begin{figure}
\vspace{0cm}
\vbox{\psfig{figure=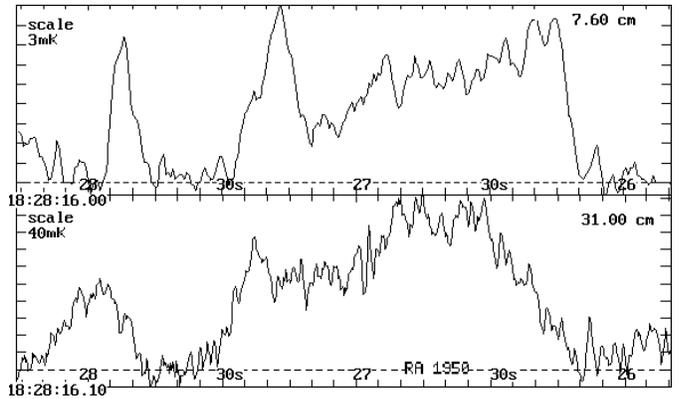,width=8.8cm}}
\vspace{0cm}
\caption[]{%
Two RATAN drift scans of the G16.2$-$2.7 at 3.9 and 0.96 GHz
at DEC1950: $-16\degr12\arcmin$. Intensity is T$_a$ in mK.
The rms is equal to 10 and 40 mJy/beam and the resolution
along RA is equal to $1\arcmin$ and $4\arcmin$ at 3.9 and 0.96 GHz,
respectively.
}
\label{scan}
\end{figure}

The sources in the field of this map from the NVSS source catalog
are marked by white crosses and the numbers around them annotate the
fractional polarization in per cent.
In Table \ref{src} a list of the NVSS sources with detectable
linear polarization is given.
This table is obtained by the {\it select}
program in the astrophysical catalogs data base CATS (Verkhodanov el al. \cite{CATS}).
We have reduced it to six columns: number, coordinates (B1950), flux density,
polarized intensity or power of polarization in per cent;
the last column indicates whether the source is a part of the SNR or not.
It is remarkable that details (included in the NVSS source catalog) of
bright western and eastern arcs of the shell are highly polarized (6--20\%
from the last column in Table \ref{src}).
Probably two relatively bright sources (14,15) are background extragalactic
ones. The weak filament-like weak source (11) in the center is highly polarized
(p=43\%) and it might be a detail of the radio shell
with a highly ordered magnetic field. The NVSS data show that the bright
eastern arc of G03.8-0.3 is also polarized to 10\% at 1.4 GHz.

\begin{figure}
\vspace{0cm}
\vbox{\psfig{figure=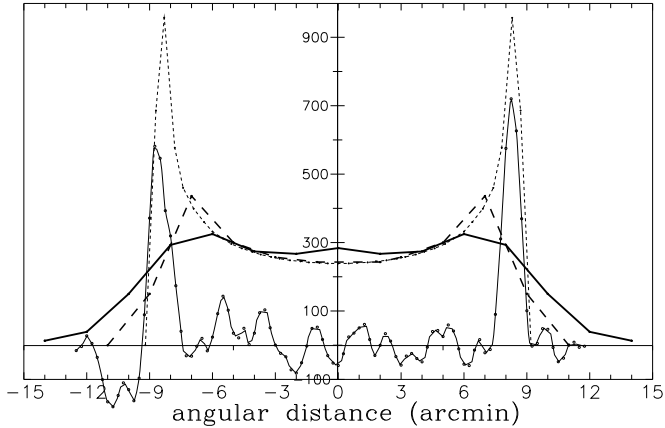,width=8.8cm}}
\vspace{0cm}
\caption[]{%
The  radial profiles of brightness:
{\it thin  solid line}  --  the real East--West one at 1.4 GHz (NVSS);
{\it thick solid line}  --  the mean East--West one at 2.7 GHz (RFRR);
{\it thin  dashed line}  --  the model profile of the hollow spherical shell:
R$_{\mathrm{max}}$ = $9\farcm2$, $\Delta$R=$1\arcmin$;
{\it thick dashed line}  --  the smoothed model with $4\farcm3$-beam
of the survey at 2.7 GHz.
}
\label{mod}
\end{figure}

The source G16.2$-$2.7 is visible in the Effelsberg surveys maps at
1.4 and 2.7 GHz (Reich et al. \cite{E21}, Reich et al. \cite{E11},
here RRF and RFRR) as a extended source, which
has not been included in the catalogs of these surveys because
its apparent size is over 16\arcmin.
We cut the maps from the original maps with a ``postage stamps'' procedure
at Max-Planck-Institut fuer Radioastronomie (MPIfR) web-site, which allowed us
to cut small images from a single survey map.
In Fig.\ref{ir} the ISSA (IRAS) $1\fdg1\times1\fdg1$ grey scale and
white contour map at 60 $\mu$m (Beichman et al. \cite{IRAS}) is superposed on
the black contour plot of the 2.7 GHz intensity map.
There is no clear relation between radio and infrared radiation.
Nevertheless, we have estimated the infrared flux to be approximately
250 and 700 Jy at 60 and 100 $\mu$m respectively within the region of
the radio shell. These fluxes are higher than $\sim100$ times the radio ones.

We used the radial profiles from the NVSS map and the 2.7GHz map
to fit the spatial parameters of the shell SNR. Model profiles of
a optically thin synchrotron hollow circular shell with an outer
diameter D, width $\Delta$R and a random magnetic field
was discussed by Rosenberg (\cite{Cas}) for Cas~A.
The best fit with the real radial profiles
at 1.4 and 2.7 GHz gives D=$18.4\arcmin$ and $\Delta$R=$1\arcmin$.
This initial radial profile
was convolved with the $4\farcm3$-beam of the Effelsberg survey at 2.7 GHz.
In Fig.\ref{ir} we compare this smoothed model profile with the mean
one from map at 2.7 GHz. We see that these profiles are very similar
within uncertainties of the background level and smoothing effects.

The low value of the flux at 1.4 GHz obtained from the NVSS image could be
explained with the above model. If a background
with a filtering window of $10\arcmin$ is subtracted from the image with
such a radial profile, then the total flux reduced to only 35\% of the
initial value becomes visible mainly in the bright limb.
Therefore, probably the real flux of the source is about 2 Jy,
which roughly coincides with the obtained spectrum.
Thus SNR G16.2$-$2.7 is likely a hollow, almost spherical
shell of optically thin smoothly distributed radio emission.

Unfortunately there are not any extended or filamental details in the
Digital Sky Survey II image of the SNR area that could be recognized as
a optical counterpart.

\section{Radio spectrum}

The source G16.2$-$2.7 was detected in the Galactic plane survey
(Trushkin, \cite{Sur96,Sur98}) at 0.96 and 3.9 GHz, then we observed
this sources in May 1999 with the RATAN-600 radio telescope. The transit
observations were carried out with the North sector at 0.96, 2.3, and 3.9 GHz
of the continuum radiometric complex of RATAN-600 in the upper culmination
of the source. The angular resolution (HPBW) in Azimuth $\times$ Elevation is
$4\arcmin\times75\arcmin$ and $1\arcmin\times40\arcmin$
at 0.96 and 3.9 GHz, respectively.

Also we used the images in the FITS format from the Effelsberg Galactic
plane surveys at 1.4 and 2.7 GHz to estimate the flux densities from
G16.2$-$2.7. In  the package ``Skyview'' (Ebert et al. \cite{sky})
we calculated the integral flux densities in a defined elliptical area of
the images.
The linear background was subtracted using the pixels of nearby regions
around G16.2$-$2.7.

\begin{table}
\caption{Measured flux densities from G16.2$-$2.7}
\begin{tabular}{llll}
\hline
  $\nu$\,(GHz) &S$_\nu$\,(Jy)&$\Delta$S$_\nu$\,(Jy)& References \\
\hline
      0.96        &  2.10     &    0.15            & this paper \\
      1.4         &  1.80     &    0.10            & map from RRF \\
      2.3         &  1.35     &    0.10            & this paper \\
      2.7         &  1.25     &    0.10            & map from RFRR \\
      3.9         &  1.03     &    0.10            & this paper \\
\hline
\end{tabular}
\label{flux}
\end{table}

   \begin{figure}
      \vspace{0cm}
\vbox{\psfig{figure=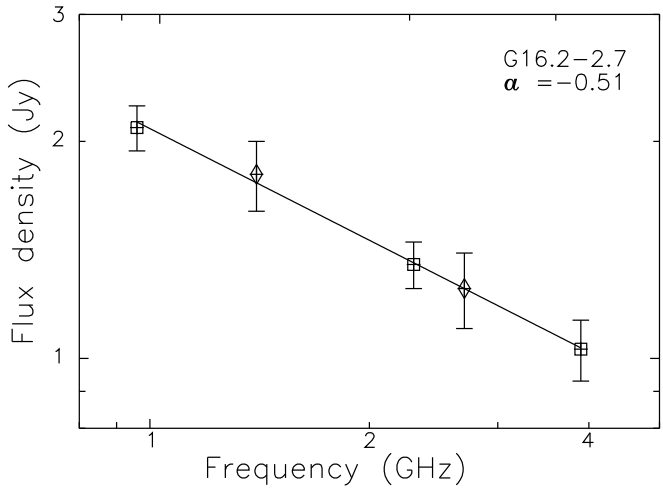,width=8.8cm}}
      \vspace{0cm}
      \caption[]{Radio spectrum of the new SNR,
    the Effelsberg surveys data at 1.4 and 2.7 GHz are marked by boxes,
    the RATAN data at 0.96, 2.3 and 3.9 GHz are marked by diamonds.
	      }
	 \label{sp}
   \end{figure}
%

In Table \ref{flux} we summarize the flux density measurements,
where $\Delta$S$_\nu$ means $1\sigma$ error.
In Fig.\ref{sp} radio spectrum of G16.2$-$2.7 is plotted.
A power law non-thermal spectrum is a good fit:
S$_\nu$[Jy]=($2.08\pm0.12$)$\,\nu^{-0.51\pm0.1}_{[\mathrm{GHz}]}$.
Based on the spectra of nearly 200 Galactic SNRs,
Trushkin (\cite{SP}) shows that the mean spectral index for the total sample
of the 200 Galactic SNRs is equal to $-0.50\pm0.1$ at high frequencies.

We have tried to roughly estimate the distance and diameter of the SNR from
the $\Sigma-D$ relation. In order to avoid uncertainties in  calibrating
the Galactic SNRs distances, we took the relation, based on the
sample of SNRs of nearby galaxies (Huang et al. \cite{Huang}):
$\Sigma_{8.4\mathrm{GHz}}$(W\,Hz$^{-1}$m$^{-2}$sr$^{-1}$) = $4.4\times10^{-16}\,D_{\mathrm{pc}}^{-3.5\pm0.1}$.
Then for the angular size $18.4\arcmin$  and above fitting of S$_\nu$,
the distance d$\sim$13~kpc and the diameter D$\sim$64~pc.
A new refined $\Sigma-D$ relation was obtained by Case \& Bhattachatya
(\cite{S-D}) for a sample of 36 Galactic shell SNRs:
$\Sigma_{1\mathrm{GHz}}$(W\,Hz$^{-1}$m$^{-2}$sr$^{-1}$) = $2.07^{+3.10}_{-1.24}\times10^{-17}\,D_{\mathrm{pc}}^{-2.38\pm0.26}$.
For $\Sigma$(1GHz)$ = (1\pm0.1)10^{-21}$W\,Hz$^{-1}$m$^{-2}$sr$^{-1}$
it gives D$=35^{+10}_{-5}$~pc and d=$6.5^{+2.0}_{-1.0}$~kpc
Then the distance from the Galactic plane z = 300$^{+75}_{-40}$ pc.

\section{Conclusions}

The new previously unidentified shell supernova remnant
G16.2$-$2.7 has been discovered in the First Galactic quadrant.

The radio map at 1.4 GHz from NVSS maps has been plotted and the SNR
has been clearly shown to be a barrel-shaped or bilateral SNR with
the axis of symmetry aligned closely with the Galactic plane.

Angular diameter of 17\arcmin\ of a circular shell has been fitted
to peaks of brightness on its limb.
Using a simple model of a spherical hollow shell, a outer diameter
D$_{\mathrm{max}}$ = $18\farcm4$ and width $\Delta$R=$1\arcmin$
are well fitted to the NVSS and the Effelsberg survey data.

The spectrum based RATAN-600 and Effelsberg surveys at cm wavelengths
is fitted by a power law with a spectral index $\alpha=-0.51\pm0.10$.

The polarization data from the NVSS show that the bright arcs
of the SNR are linearly polarized to 6--20\,\%,
indicating a synchrotron radiation and an ordered magnetic field in the shell.

\begin{acknowledgements}
      Part of this work is supported by the Russian ``Astronomy'' program
      project N1.3.2. I thank Booth Hartley (IPAC Caltech)
      for a good package, ``Skyview'' ver.3.3 for Linux, Dr. E. F\"{u}rst
      (Max-Planck-Institut f\"{u}r Radioastronomie) for FITS maps of the
      Galactic plane surveys and NRAO for FITS maps of the NVSS survey.
      I am very grateful to an anonymous referee for helpful suggestions.
\end{acknowledgements}

\end{document}